\documentclass[aps,prd,amsmath,amssymb,twocolumn,superscriptaddress,preprintnumbers,nofootinbib]{revtex4-1}
\usepackage[utf8]{inputenc} 
\usepackage{graphicx}% Include figure files
\usepackage{dcolumn}% Align changzhu.wu@mailbox.tu-dresden.de] columns on decimal point
\usepackage{bm}
\usepackage[bookmarks, pagebackref=false]{hyperref}
\usepackage[nameinlink]{cleveref}
\usepackage[usenames,dvipsnames]{xcolor}
\usepackage[normalem]{ulem}
\definecolor{rossoCP3}{cmyk}{0,.88,.77,.40}
\definecolor{blaa}{rgb}{0.2,0.2,0.6}
\hypersetup{colorlinks, 
	bookmarksopen, 
	bookmarksnumbered,
	citecolor=blaa, 		%color of links to bibliography
	linkcolor=rossoCP3,	%color of internal links
	urlcolor=rossoCP3,			%color of external links 
}

\crefname{section}{Sec.\!}{Secs.\!}
\crefname{figure}{Fig.\!}{Figs.\!}
\crefname{equation}{Eq.\!}{Eqs.\!}
\crefname{table}{Tab.\!}{Tabs.\!}
\crefname{appendix}{App.\!}{Apps.\!}

\usepackage{natbib}
\usepackage{slashed}
\newcolumntype{x}[1]{>{\centering\arraybackslash\hspace{0pt}}p{#1}}

\graphicspath{{./figs/}}

\begin{document}
 
\title{ \LARGE  \color{rossoCP3} Testing the dark SU(N) Yang-Mills theory Confined Landscape:  From the Lattice to Gravitational Waves}
\author{Wei-Chih {\sc Huang}}
\thanks{{\scriptsize Email}: \href{mailto:huang@cp3.sdu.dk}{huang@cp3.sdu.dk}; {\scriptsize ORCID}: \href{https://orcid.org/0000-0001-7939-3246}{0000-0001-7939-3246}}
\affiliation{CP$^3$-Origins,  University of Southern Denmark, Campusvej 55, 5230 Odense M, Denmark}

\author{Manuel {\sc Reichert}}
\thanks{{\scriptsize Email}: \href{mailto:m.reichert@sussex.ac.uk}{m.reichert@sussex.ac.uk}; {\scriptsize ORCID}: \href{https://orcid.org/0000-0003-0736-5726}{ 0000-0003-0736-5726}}
\affiliation{Department  of  Physics  and  Astronomy,  University  of  Sussex,  Brighton,  BN1  9QH,  U.K.}

\author{Francesco {\sc Sannino}}
\thanks{{\scriptsize Email}: \href{mailto:sannino@cp3.sdu.dk}{sannino@cp3.sdu.dk}; {\scriptsize ORCID}: \href{https://orcid.org/0000-0003-2361-5326}{ 0000-0003-2361-5326}}
\affiliation{CP$^3$-Origins,  University of Southern Denmark, Campusvej 55, 5230 Odense M, Denmark} 
\affiliation{Dipartimento di Fisica “E. Pancini”, Università di Napoli Federico II | INFN sezione di Napoli, Complesso Universitario di Monte S. Angelo Edificio 6, via Cintia, 80126 Napoli, Italy}

\author{Zhi-Wei {\sc Wang}}
\thanks{{\scriptsize Email}: \href{mailto:wang@cp3.sdu.dk}{wang@cp3.sdu.dk}; {\scriptsize ORCID}: \href{https://orcid.org/0000-0002-5602-6897}{0000-0002-5602-6897}}
\affiliation{CP$^3$-Origins,  University of Southern Denmark, Campusvej 55, 5230 Odense M, Denmark}

\begin{abstract}  
We pave the way for future gravitational-wave detection experiments, such as the Big Bang Observer and DECIGO, to constrain dark sectors made of $SU(N)$ Yang-Mills confined theories.  We go beyond the state-of-the-art by combining first principle lattice results and effective field theory approaches to infer essential information about the non-perturbative dark deconfinement phase transition driving the generation of gravitational-waves in the early universe, such as the order, duration and energy budget of the phase transition which are essential in establishing the strength of the resulting gravitational-wave signal.
\end{abstract}
 
\maketitle

\section{Introduction}
The null search results for dark matter (DM) via direct detection and colliders suggest that it is likely that DM resides in a hidden sector which couples weakly to the Standard Model (SM) \cite{Strassler:2006im,Cheung:2007ut,Hambye:2008bq,Feng:2009mn,Cohen:2010kn,Foot:2014uba,Bertone:2016nfn}. Yet, very little is known about the dark side of the Universe and it is therefore highly desirable to be able to test the immense landscape of available dark/hidden sectors. Here we concentrate on the well-motivated scenario that the dark side features composite sectors made by non-abelian Yang-Mills theories which are mainly gravitationally coupled.  These theories are physically motivated because the dynamics of the dark sector very naturally mimics the SM QCD featuring strong interactions. Furthermore, these theories are well-behaved at short distance denoted as asymptotically freedom \cite{Gross:1973id,Politzer:1973fx}, meaning that the theories are, per se, ultraviolet complete before coupling to gravity, and they do not introduce new types of hierarchies beyond the SM one. The latter means that the theories are stable against quantum corrections. This dynamics has been widely implemented to DM models \cite{DelNobile:2011je, Hietanen:2013fya, Cline:2016nab, Cacciapaglia:2020kgq, Dondi:2019olm, Ge:2019voa, Beylin:2019gtw, Yamanaka:2019yek, Yamanaka:2019aeq}.

These type of theories are unfortunately inaccessible to current colliders or direct searches, limiting our ability to test them and therefore pin down the model underlying the dark sector. Here we propose to investigate the dynamics of this hidden sector via the detection of gravitational waves (GWs) in a {\it model-independent} fashion. We make as few assumptions as possible on the specific extensions of the SM, and assume the minimal interaction of gravity between the SM and new strongly-coupled sectors. Intriguingly, GWs provide a unique window for detecting the dark deconfinement phase transition\footnote{Confinement occurs at sufficiently low temperatures when gluons form composite states known as glue balls. At high temperature the theory deconfines the gluons by melting the composite states.  Therefore we indicate by {\it Dark} confinement-deconfinement to the expected phase transition as function of the temperature occurring during the evolution of the universe and taking place in the hidden sector.}.

To set the stage, we assume that the dark landscape is constituted by $n$ copies of $SU(N)$ Yang-Mills confined theories for a given confinement scale. To determine the relevant physical information in strongly-coupled theories we often require lattice simulations and/or effective field theory approaches. We adopt state-of-the-art results of lattice simulations \cite{Lucini:2005vg, Panero:2009tv} combined with well-defined effective approaches \cite{Pisarski:2000eq, Pisarski:2001pe, Pisarski:2002ji, Sannino:2002wb} to precisely pin down the nonperturbative physics involved in the (dark) deconfinement phase transition as functions of the temperature and number of dark colours. For the effective description we marry the Polyakov loop action \cite{Pisarski:2000eq, Pisarski:2001pe, Pisarski:2002ji, Sannino:2002wb} with lattice simulations. The $N=3$ case was extensively investigated in the literature \cite{Ratti:2005jh, Fukushima:2013rx, Fukushima:2017csk} while here we go beyond the state-of-the-art by incorporating the lattice results \cite{Panero:2009tv} at the effective action level for $N=4,\, 5,\, 6$, and $8$. These cases are phenomenologically motivated as they arise in a variety of Grand Unified Theories and composite models. Our work thus covers a wide range of theories. We carefully analyse the dark phase transition for the first few numbers of colours and then generalise it to arbitrarily large numbers. This allows us to acquire an unprecedented eagle view on the dynamics involved in phase transitions of dark composite models by generalising the results to arbitrary numbers of colours. In our work, we go beyond the state-of-the-art by connecting different research fields from (astro-)particle physics over first principle numerical simulations to GW astronomy. It underscores the necessity of orchestrated plans and efforts to unravel the enigma on the nature of DM.

We investigate the GW generation triggered by the dark confinement phase transition discovering, for our generic setup, that: (i) The strength parameter $\alpha$, related to the energy budget of the phase transition, takes values around $\alpha \approx 1/3$, while the parameter $\beta$, that measures the inverse duration of the phase transition, assumes values of the order of $10^4-10^5$ in units of the Hubble time. (ii) The GW signal emerging from sound waves dominates over the bubble collision and turbulence due to the impact of the friction term \cite{Bodeker:2009qy, Bodeker:2017cim} related to the bubble-wall velocity. (iii) The strength of the induced GW signal is nearly independent of the number of colours for $N\geq 6$. That is because the strength depends on the jump in the entropy across the deconfinement phase transition per degree of freedom rather than on the overall jump in entropy, which is inevitably proportional to $N^2$. The strength of the GW signal culminates at the case of $SU(6)$ and then gradually decreases with an increasing number of dark colours. Also the peak frequency increases with the number of colours.

\begin{figure*}[t!]
	\centering  
	\includegraphics[width=\linewidth]{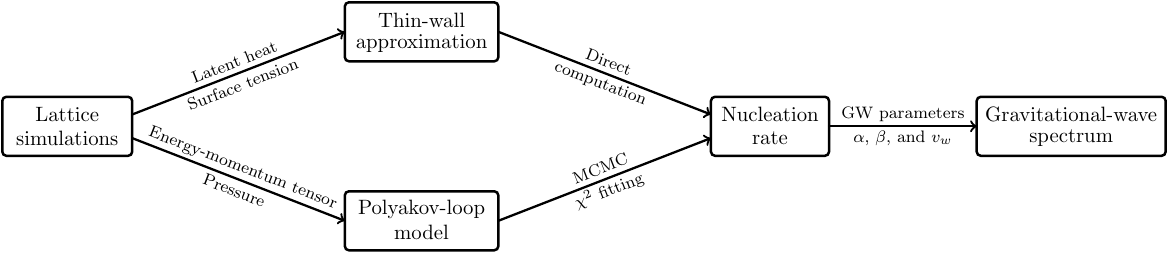}
	\caption{Schematic summary of our work flow from lattice simulations to gravitational waves. The lattice results for the latent heat and the surface tension are taken from \cite{Lucini:2005vg}, while the ones for the energy-momentum tensor and the pressure are taken from \cite{Panero:2009tv}.}
	\label{fig:work-scheme}
\end{figure*}

The bubble profile and the nucleation rate can be also directly computed in the thin-wall approximation, which allows us to have an independent check of our results from the effective Polyakov loop model. The analysis procedure is neatly summarised by the flow chart in \cref{fig:work-scheme}. In the thin-wall approximation, the nucleation rate is directly obtained from the latent heat and the surface tension, which have been computed with lattice simulations \cite{Lucini:2005vg, Panero:2009tv}. The results from both methods are in qualitative agreement, reinforcing the validity and consistency of our work.

 %%%%%
%%%%%

We compute the constraints on the dark confined landscape from the next generation of GW observatories including LISA \cite{Audley:2017drz, Baker:2019nia,LISAdocument}, the Big Bang Observer (BBO) \cite{Crowder:2005nr, Corbin:2005ny, Harry:2006fi,Thrane:2013oya, Yagi:2011wg}, DECIGO \cite{Seto:2001qf,Yagi:2011wg, Kawamura:2006up,Isoyama:2018rjb}, the Einstein Telescope (ET) \cite{Punturo:2010zz, Hild:2010id, Sathyaprakash:2012jk, Maggiore:2019uih}, and the Cosmic Explorer (CE) \cite{Evans:2016mbw, Reitze:2019iox}. The signal-to-noise ratio of all experiments over the dark confinement scales from the MeV to PeV scale is shown in \cref{fig:SNR}. Intriguingly, for confinement temperatures from one to a few hundred GeV, the full range of theories will be independently tested by BBO and DECIGO. They could either constrain such dark dynamics or more excitingly detect signals.

This work constitutes a stepping stone towards embarking in a careful analysis of dark sectors featuring both dark gluons and quarks \footnote{For any strongly coupled (composite) theory, the pure gluon dynamics are the key-ingredients and backbones. Thus, our work paves the road to study more elaborate models. For earlier analyses of the chiral phase transition see \cite{Jarvinen:2009mh, Schwaller:2015tja, Chen:2017cyc, Helmboldt:2019pan, Agashe:2019lhy, Bigazzi:2020avc}.}. In this case, the relevant phase transitions include the dark deconfinement and the dark chiral phase transition. We can take into account these transitions by extending the current work to properly marrying lattice data with the appropriate effective actions introduced first in \cite{Mocsy:2003tr, Mocsy:2003qw}.

\section{The Polyakov Loop Model}
\label{sec:Polyakov_model}
\subsection{Polyakov Loop}
In this work, we consider $SU(N)$ Yang-Mills theory at finite temperature $T$. The dynamics is purely gluonic and no fermions are involved. Following 't Hooft \cite{tHooft:1977nqb, tHooft:1979rtg}, in any $SU(N)$ gauge theory, a global $Z_N$ symmetry, called the central symmetry, naturally emerges from the associated local gauge symmetry. It is possible to construct a number of gauge invariant operators charged under this global $Z_N$ symmetry. Among them, the most notable one is the Polyakov loop,
\begin{align}
{\ell}\left(x\right)=\frac{1}{N}{\rm Tr}[{\bf L}]\,,
\label{eq:Polyakov_Loop}
\end{align}
where 
\begin{align}
{\bf L}={\mathcal P}\exp\!\left[i\,g\int_{0}^{1/T} \!\! A_{0}(x,\tau)\,\mathrm d\tau\right],
\end{align}
is the thermal Wilson line, $\cal P$ denotes the path ordering, $g$ is the $SU(N)$ gauge coupling, and $A_0$ is the vector potential in the time direction. The symbols $x$ and $\tau$ denote the three spatial dimensions and the Euclidean time, respectively. The Polyakov loop can be transformed under the $Z_N$ symmetry
\begin{align}
\ell&\rightarrow e^{i\phi}\ell\,,
&
\phi&=\frac{2\pi j}{N}\,,
&
j&=0,1,\ldots,(N-1)\,.
\label{eq:Zn_transformation}
\end{align}
The phase $\phi$ shows the discrete symmetry $Z_N$. From \eqref{eq:Zn_transformation}, it is clear that $\ell$ is real when $N=2$ and otherwise $\ell$ is complex. An important feature of the Polyakov loop is that its expectation value vanishes below the critical temperature $T_c$, i.e.\ $\langle\ell\rangle_{T<T_c}=0$, while it possesses a finite expectation value above the critical temperature, i.e.\ $\langle\ell\rangle_{T>T_c}>0$ . In fact, at very high temperature, the allowed vacua exhibit a $N$-fold degeneracy and we have
\begin{align}
\langle \ell\rangle&=\exp\!\left(i\frac{2\pi j}{N}\right)\ell_0\,,
&
j&=0,1,\ldots,(N-1)\,,
\end{align}
where $\ell_0$ is defined to be real and $\ell_0\rightarrow 1$ as $T\rightarrow\infty$. Thus, the Polyakov loop is a suitable order parameter in the finite temperature phase transition of the $SU(N)$ gauge theory.

\subsection{Effective Potential of the Polyakov Loop Model}
The Polyakov Loop Model (PLM) was proposed by Pisarski in \cite{Pisarski:2000eq, Pisarski:2001pe} as an effective field theory to describe the confinement-deconfinement phase transition of the $SU(N)$ gauge theory. The Polyakov loop \eqref{eq:Polyakov_Loop} plays the role of an order parameter. The simplest effective potential preserving the $Z_N$ symmetry is given by
\begin{subequations}
\label{eq:PLM_full}
\begin{align}
V_{\rm{PLM}}=T^4\left(-\frac{b_2(T)}{2}|\ell|^2+b_4|\ell|^4-b_3\!\left(\ell^{N}+\ell^{*N}\right)\right),
\label{eq:PLM_potential}
\end{align}
where 
\begin{align}
b_2(T)=a_0+a_1\!\left(\frac{T_c}{T}\right)\!+a_2\!\left(\frac{T_c}{T}\right)^{\!2}\!+a_3\!\left(\frac{T_c}{T}\right)^{\!3}\!+a_4\!\left(\frac{T_c}{T}\right)^{\!4}\!\!.
\label{eq:b2}
\end{align}
\end{subequations}
We have chosen the coefficients $b_3$ and $b_4$  to be temperature independent following the treatment in \cite{Ratti:2005jh, Fukushima:2017csk}, which studied the $SU(3)$ case, and also neglected higher orders in $\lvert\ell\rvert$ in \eqref{eq:PLM_potential}. Note that there is no $a_4$ term in the parameterize of $b_2(T)$ in \eqref{eq:b2} in \cite{Ratti:2005jh, Fukushima:2017csk} while we find it can improve the chi-square fitting discussed below. The $a_2$ term in \eqref{eq:b2} has the physics meaning of the ``fuzzy bag" term in the ``fuzzy bag" model\footnote{In the Fuzzy Bag model the pressure as a function of temperature $p(T)$ is written as $p(T)=f_\text{pert}T^4-B_\text{fuzzy}T^2-B_{\rm{MIT}}$ where $f_\text{pert}$ denotes the perturbative contributions, $B_\text{fuzzy}$ is the ``fuzzy bag" term and $B_{\rm{MIT}}$ is the term associated with the usual MIT bag model.} proposed in \cite{Pisarski:2006yk} as a generalization of the famous MIT bag model \cite{Chodos:1974je}. On the other hand the $a_4$ term actually captures the low temperature information and is equivalent to the ${\mathcal P}[\ell]$ contribution\footnote{In \cite{Sannino:2002wb}, it was proposed that the total effective potential $V[\ell]$ can be written as $V[\ell]=T^4{\cal V}[\ell]+\frac{\Lambda}{e}\cal{P}[\ell]$ where $\Lambda$ is the confining scale, ${\cal V}[\ell]=a_1\ell^2+a_2\ell^4$ and ${\cal P}[\ell]=b_1\ell^2$.} in the model proposed in \cite{Sannino:2002wb}. 

Note that the above PLM potential \eqref{eq:PLM_potential} is the minimal case since we have only considered the Polyakov loop with charge one. For higher charge cases, say charge two cases, the effective potential will be similar to a multi-scalar fields Higgs portal model (see e.g.~\cite{Pisarski:2002ji}). However, in the special case where the higher charge Polyakov loop is heavy and can be integrated out, the low energy effective field theory shares a similar form as the current setting of the PLM potential in \eqref{eq:PLM_full}.

With the set-up of the PLM effective potential using \eqref{eq:PLM_full}, we study the $SU(N)$ Yang-Mills theory with $N=3,\,4,\,5,\,6$, and $8$. By choosing these numbers of colours, we can take the advantage of the existing lattice data \cite{Panero:2009tv}. In the following, we explicitly list the PLM potential corresponding to the number of colours. Extra terms are added to some of the cases such that the potential is bounded from below (the $SU(4)$ case) or the fit to the data is decent -- chi-squared per degree of freedom is around or below one (the $SU(6),\, SU(8)$ cases). 

For the $SU(3)$ and $SU(5)$ cases, the PLM potential is exactly given by the formula \eqref{eq:PLM_potential} with $N=3, \, 5$. For the $SU(3)$ case, there is also an alternative logarithmic parameterization, see e.g.~\cite{Fukushima:2017csk, Roessner:2006xn}, given by
\begin{align}
\label{PLM_SU3_Log}
V_{\rm{PLM}}^{(3)}&=T^4\bigg(\!-\frac{a(T)}{2}|\ell|^2 \\
&\quad +b(T)\ln\!\left[1-6|\ell|^2+4(\ell^{*3}+\ell^{3})-3|\ell|^4 \right]\!\bigg),\notag
\end{align}
with
\begin{align}
a(T) & =a_0+a_1\!\left(\frac{T_c}{T}\right)+a_2\!\left(\frac{T_c}{T}\right)^{\!2}+a_3\!\left(\frac{T_c}{T}\right)^{\!3}\!, \notag \\
b(T) & = b_3\!\left(\frac{T_c}{T}\right)^{\!3}\!. 
\end{align}
The coefficients inside the logarithm are determined by the Haar measure for which the explicit form for $SU(N)$ with $N>3$ is unknown. Thus, we do not have a logarithmic parameterization for $N>3$.

For the $SU(4)$ case, the PLM potential is more subtle since the $b_3$ term is given by $\ell^{4}+\ell^{*4}$ and thus of the same order as the $b_4$ term. As consequence, their effects are indistinguishable for real values of $\langle\ell\rangle$ and we have to introduce an $|\ell|^6$ term to properly parameterize the lattice results \cite{Panero:2009tv, Lucini:2005vg}. Thus, the PLM potential for $SU(4)$ is given by
\begin{align}
V_{\rm{PLM}}^{(4)}=T^4\!\left( - \frac{b_2(T)}{2}|\ell|^2+b_4|\ell|^4+b_6|\ell|^6\right),
\label{eq:PLM_SU4}
\end{align}
where $b_2(T)$ is given by \eqref{eq:b2}. For the $SU(6)$ and $SU(8)$ cases, the PLM potentials are parametrized in the same way as
\begin{align}
V_{\rm{PLM}}^{(6,8)}=T^4\!\left( - \frac{b_2 (T)}{2}|\ell|^2+b_4|\ell|^4 +b_6|\ell|^6 +b_8|\ell|^8 \right),
\label{eq:PLM_SU6}
\end{align}
where $b_2(T)$ is again given by \eqref{eq:b2}. We emphasise that we could include higher-order terms such as $|\ell|^8$ in the potentials for $SU(3)$ and $SU(4)$ (\eqref{eq:PLM_potential} and \eqref{eq:PLM_SU4}) but they would not improve the fit on the Lattice data and the respective coefficient $b_8$ would be strongly suppressed.

 %%%%%
\begin{figure}[t!]
	\centering
	\includegraphics[width=\linewidth]{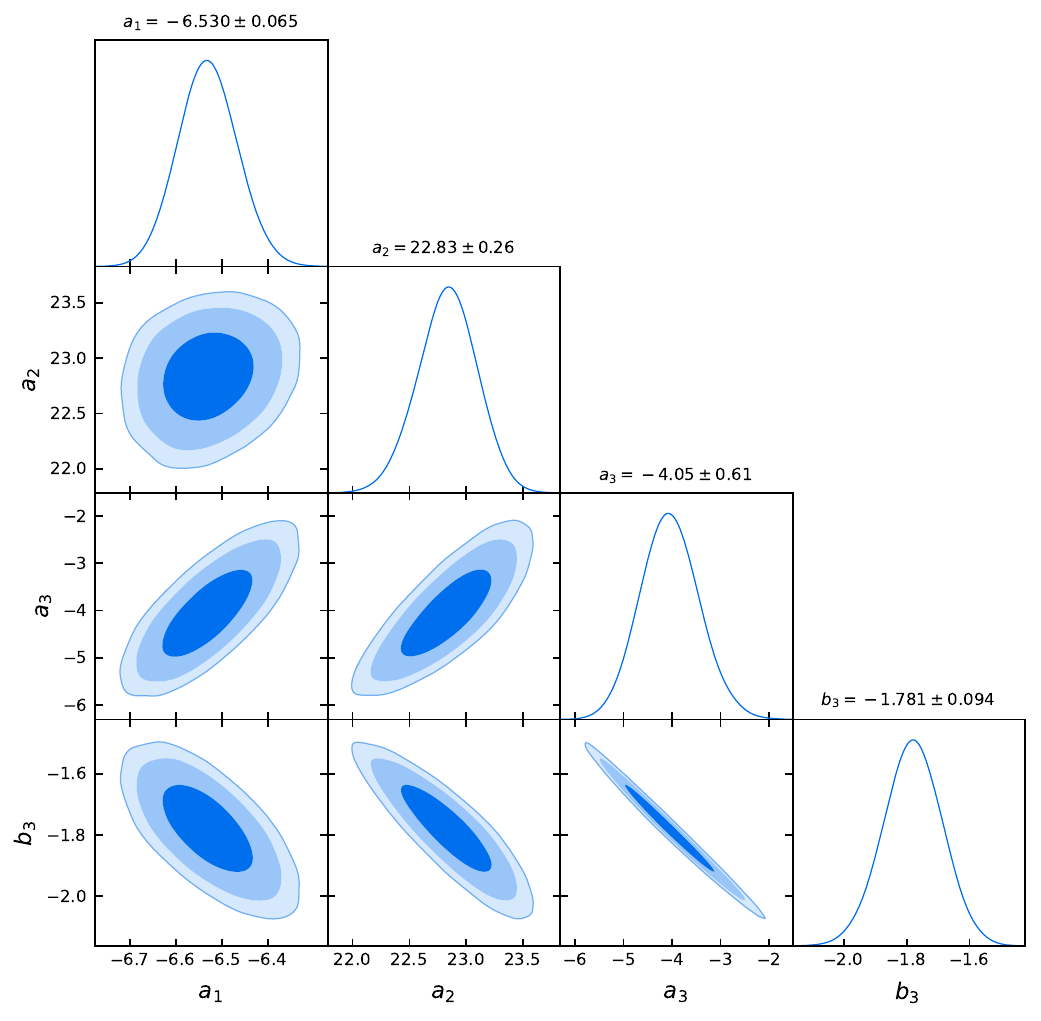}
	\caption{Preferred regions for the log parametrization of the $SU(3)$ case where the three areas correspond to the confidence level
		of $68 \%$, $95 \%$, and $99 \%$, respectively.}
	\label{fig:SU3_tri}
\end{figure}
%%%%%

 %%%%%
\begin{figure*}[t!]
	\centering  
	\includegraphics[width=\linewidth]{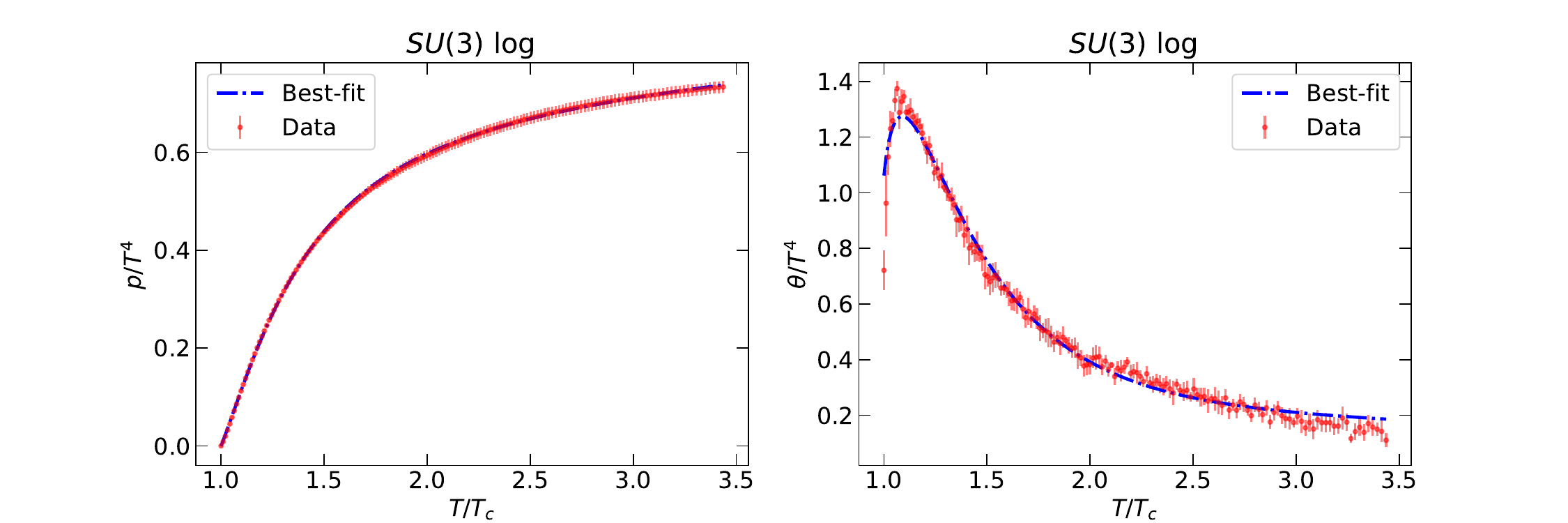}
	\caption{Comparison between the data~(red) and best-fit curves~(blue) for the pressure~(left panel) and trace of the energy momentum tensor~(right panel). Note that we normalize both $p$ and $\theta$ with respect to the SB limit as done in \cite{Panero:2009tv}. }
	\label{fig:SU3_best}
\end{figure*}
%%%%%

\subsection{Fitting the PLM potential to lattice data}
\label{sec:fitting}
With the explicit PLM effective potential for different colours, we are now able to determine the parameters $b_i$ by fitting the potential to the lattice results in \cite{Panero:2009tv}. The thermodynamical observables measured on the lattice are the pressure $p$, the energy density $e$ and the trace of the energy-momentum tensor $\theta$\footnote{Our $\theta$ is defined the same as $\Delta$ in paper \cite{Panero:2009tv}.} and the entropy density $s$. The lattice simulations compute the difference between the finite temperature expectation value and the zero temperature one. The energy density $e$ and the entropy density $s$ can be written as linear combinations of the pressure $p$ and the trace of energy-momentum tensor~$\theta$
\begin{align}
e&=\theta+3p\,,
&
s&=\frac{\theta+4p}{T}\,.
\end{align}
Thus we only use the lattice data of $\theta$ and $p$ from \cite{Panero:2009tv} to determine the coefficients of $a_i$ and $b_i$ in the above PLM potential setting. We only have access to the statistical uncertainties and therefore we inflated them by a factor of two to mimic the effect of the systematic uncertainties.

During the chi-square~($\chi^2$) analysis, we impose the Stefan-Boltzmann~(SB) limit: $| \ell | \to 1$ for $T \to \infty$ and $p/T^4 \vert_{T\to \infty} \to 1.21\cdot(N^2-1)\cdot\pi^2/45$ \cite{Panero:2009tv}, which provides two constraints on the parameters of the polynomial parameterizations but only one constraint for the logarithmic case. The above guarantees that the pressure approaches the ideal gas law at infinite temperature. Additionally, the parameters of $a_i$ and $b_i$ need to fulfil the constraint that $T_c$, which is a priori only a parameter in \eqref{eq:PLM_full}, is indeed the critical temperature, i.e., the temperature at which two minima are degenerate.

We employ the \texttt{Python} package \texttt{emcee} \cite{ForemanMackey:2012ig}, which is based on Affine Invariant Markov chain Monte Carlo~(MCMC) Ensemble sampler, to find favourable regions of the parameter space. In \cref{fig:SU3_tri}, with the help of the analysis tool for MCMC samples, \texttt{GetDist}~\cite{Lewis:2019xzd}, we display the best-fit regions of the $SU(3)$ log case where $a_0$ is fixed by the SB limit with $a_0 = 4.26$. In \cref{fig:SU3_best}, we demonstrate how well the best-fit point, with a reduced $ \chi^2 = 0.70$, can fit both, $p$ and $\theta$. We present the best-fit values of the potential parameters for the colours $N=3,\,4,\,5,\,6$, and $8$ in \cref{tab:best-fit}.
 
 %%%%
 \begin{table}[b]
 	\centering
  	\caption{The parameters for the best-fit points.}
  	\begin{tabular}{|x{.8cm}|x{1.13cm}|x{1.13cm}|x{1.13cm}|x{1.13cm}|x{1.13cm}|x{1.13cm}|}
 	\hline
 	$N$ & 3  & $3\;\log$& 4 & 5 & 6 & 8 \\
 	\hline
 	$a_0$ & 3.78 & 4.26 & 9.58 & 11.4 & 11.2 & 20.1 \\
 	\hline
 	$a_1$ & -5.48 & -6.53 & -8.81  & -12.2 & -29.1 & -52.3 \\
 	\hline
 	$a_2$ & 8.47 & 22.8 & 10.1 & 4.41 & 67.1 & 121 \\
 	\hline
 	$a_3$ & -9.47 & -4.10 & -12.2 & -0.148 & -95.6 & -172 \\
 	\hline
 	$a_4$ & 0.222 & & 0.475 & -8.29 & 32.9 & 59.2 \\
 	\hline
 	$b_3$ & 2.36 & -1.77 & & -7.03 & & \\
 	\hline
 	$b_4$ & 4.49 & & -2.37 & -14.7 & -29.3 & -52.8  \\
 	\hline
 	$b_6$ & & & 3.18 & & 38.2 & 68.8 \\
 	\hline	
 	$b_8$ & & & & & -12.6 & -22.7 \\
 	\hline											
 \end{tabular}
 	\label{tab:best-fit}
 \end{table}
%%%%

\section{First-order Phase Transition and Gravitational Waves}
\label{sec:PhaseTransition}
In this section, we discuss the order of the confinement-deconfinement phase transition and the resulting GW signal. We start with a brief review of the bubble nucleation process and the computation of the GW parameters $\alpha$ (strength parameter), $\beta$ (inverse duration time), and $v_w$ (bubble-wall velocity). Then, we discuss the analytic results obtained from the thin-wall approximation and compare with the PLM fitting results. Remarkably, the analytic results from the thin-wall approximations show interesting patterns that are consistent with those of the fitting to the lattice results. For reviews on GWs from first-order phase transitions see, e.g., \cite{Cai:2017cbj, Weir:2017wfa, Caprini:2018mtu, Caprini:2019egz, Wang:2020jrd, Hindmarsh:2020hop}.

\subsection{Bubble nucleation}
\label{sec:bubble-nucleation}
In this section, we briefly review the generic picture of bubble nucleation processes where some subtleties related to our models are emphasized. 

The conventional picture of a first-order phase transition is that, as the universe cools down, a second minimum with a non-zero vacuum expectation value (broken phase) develops at a critical temperature. This triggers the tunnelling from the false vacuum (unbroken phase) to the stable vacuum (broken phase) below the critical temperature. In our model, this picture is reversed -- in a sense, as the universe cools down, the tunnelling occurs from the broken phase (deconfinement phase) to the unbroken phase (confinement phase). The underlying reason behind this reversed phenomenon is that the discrete symmetry $Z_N$ is broken in the deconfinement phase at high temperature while it is preserved at the confinement phase at low temperature. 

\begin{figure}[t!]
\centering
\includegraphics[width=\linewidth]{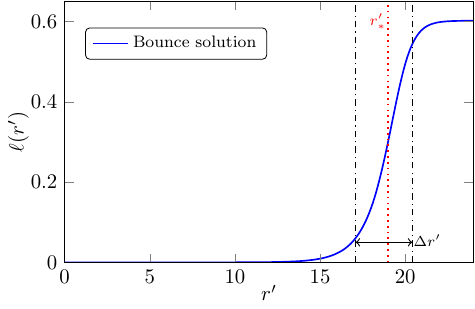}
\caption{Bounce solution in the case of $SU(4)$. The bubble radius is indicated by $r'_*$ and the wall width by $\Delta r'$. Inside of the bubble ($r'\ll r'_*$), the $Z_N$ symmetry is unbroken and $\langle\ell\rangle =0$, while outside of the bubble ($r\gg r'_*$), the $Z_N$ symmetry is broken and $\langle\ell\rangle >0$. }
\label{fig:instanton_solution}
\end{figure}

The tunnelling rate due to thermal fluctuations per unit volume as a function of the temperature from the metastable vacuum to the stable one is suppressed by the three-dimensional Euclidean action $S_3(T)$ \cite{Coleman:1977py, Callan:1977pt, Linde:1980tt, Linde:1981zj} and we have
\begin{align}
\Gamma(T)=T^4\left(\frac{S_3(T)}{2\pi T}\right)^{\!3/2} e^{-S_3(T)/T}.
\label{eq:decay_rate}
\end{align}
The three-dimensional Euclidean action reads
\begin{align}
S_3(T)=4\pi\!\int_0^\infty \!\!\mathrm dr\,r^2\!\left[\frac{1}{2}\!\left(\frac{\mathrm d\rho}{\mathrm dr}\right)^{\!2} +V_\text{eff}(\rho,T)\right],
\label{eq:Euclidean_Action_general}
\end{align}
where $\rho$ is a scalar field with the effective potential $V_\text{eff}$. The scalar field $\rho$ has mass dimension one, $\left[\rho\right]=1$, in contrast to the Polyakov loop $\ell$, which is dimensionless. Furthermore, $V(\rho,T)$ has mass dimension four. After rewriting the scalar field as $\rho=\ell\, T$ and converting the radius into a dimensionless quantity $r'=r\, T$, the action becomes
\begin{align}
S_3(T)=4\pi T \!\int_0^\infty \!\!\mathrm dr'\, r'^2\!&\left[\frac{1}{2}\left(\frac{\mathrm d\ell}{\mathrm dr'}\right)^{\!2} +V'_\text{eff}(\ell,T)\right]\,,
\label{eq:Euclidean_Action}
\end{align}
which has the same form as \eqref{eq:Euclidean_Action_general}. Here, $V'_\text{eff} (\ell,T) \equiv V_\text{eff}(\ell,T)/T^4$ is dimensionless. Keep in mind that $r'$ in the bubble-profile solution is not the physical bubble radius but the product of bubble radius and the temperature. The bubble profile (instanton solution) is obtained by solving the equation of motion of the action in \eqref{eq:Euclidean_Action}
\begin{align}
\frac{\mathrm d^2\ell(r')}{\mathrm dr'^2}+\frac{2}{r'}\frac{\mathrm d\ell(r')}{\mathrm dr'}-\frac{\partial V_\text{eff}'(\ell,T)}{\partial\ell}=0\,,
\label{eq:bounce-solution}
\end{align}
with the associated boundary conditions
\begin{align}
\frac{\mathrm d\ell(r'=0,T)}{\mathrm dr'}&=0\,,
&
\lim_{r'\rightarrow 0} \ell(r',T)&=0\,.
\end{align}
To attain the solutions, we used the method of overshooting/undershooting and employ the \texttt{Python} package \texttt{CosmoTransitions} \cite{Wainwright:2011kj}. A sample plot is shown in \cref{fig:instanton_solution} where we also indicate the thickness of the bubble wall, which we will use later. We substitute the bubble profile $\ell(r',T)$ into the three-dimensional Euclidean action \eqref{eq:Euclidean_Action} and, after integrating over $r'$, $S_3$ depends only on $T$.

%%%%%
%%%%%

\subsection{Inverse duration time of the phase transition}
\label{sec:beta}
An important parameter for the computation of the GW signal from a first-order phase transition is the inverse duration time $\beta$. For sufficiently fast phase transitions, the decay rate can be approximated by
\begin{align}
 \Gamma(T) \approx \Gamma(t_*) e^{\beta (t-t_*)}\,,
 \label{eq:approx-Gamma}
\end{align}
where $t_*$ is the characteristic time scale for the production of GWs. The inverse duration time then follows as
\begin{align}
\beta= - \frac{\mathrm d}{\mathrm dt} \frac{S_3(T)}{T}\bigg\vert_{t=t_*}\,.
\label{beta}
\end{align}
The dimensionless version is obtained by dividing with the Hubble parameter $H$
\begin{align}
\tilde \beta = \frac{\beta}{H_*}=T\frac{\mathrm d}{\mathrm dT}\frac{S_3(T)}{T}\bigg\vert_{T=T_*}\,,
\label{eq:beta-tilde}
\end{align}
where we used that $\mathrm dT/\mathrm dt = -H(T)T$. 

Note that in the above analysis, we have implicitly assumed that the temperature in the strongly coupled hidden sector (denoted as $T_d$) and the temperature in the visible sector ($T_v$) are the same (i.e.~$\xi \equiv T_d/T_v=1$). In general, these two temperatures can be different. In this case, the inverse duration is given by
\begin{align}
	\tilde \beta = \frac{\beta}{H(T_v, T_d)}=T_d\frac{\mathrm d}{\mathrm dT_d}\frac{S_3(T_d)}{T_d}\bigg\vert_{T_d=T_{d,*}}\,,
\end{align}
with Hubble parameter $H(T_v, T_d)$ given
\begin{align}
	H(T_v,T_d) \sim \sqrt{g_{*,d} T_d^4 + g_{*,v} T_v^4 } = \sqrt{g_{*,d} \xi^4 + g_{*,v}} T_v^2 \,.
\end{align}
Here, $g_{*,d}$ and $g_{*,v}$ are the effective number of relativistic degrees of freedom in the hidden and visible sector, respectively.

The phase transition temperature $T_*$ is often taken as the nucleation temperature $T_n$, which is defined as the temperature at which the rate of bubble nucleation per Hubble volume and time is approximately one, i.e.\ $\Gamma/H^4\sim \mathcal{O}(1)$. A more accurate definition is to use the percolation temperature $T_p$, which is defined as the temperature at which the probability to have the false vacuum is about $0.7$. For very fast phase transitions, as in our case, the nucleation and percolation temperature are almost identical $T_p\lesssim T_n$. However, even a small change in the temperature leads to an exponential change on the vacuum decay rate 
$\Gamma$, see \eqref{eq:approx-Gamma}, and consequently we use the percolation temperature throughout this work. We write the false-vacuum probability as~\cite{Guth:1979bh, Guth:1981uk}
\begin{align}
P(T) = e^{-I(T)}\,,
\end{align}
with the weight function \cite{Ellis:2018mja}
\begin{align}
I(T)=\frac{4\pi}{3} \int^{T_c}_T \! \!\mathrm dT'\frac{\Gamma(T')}{H(T')T'{}^4} \left( \int^{T'}_{T}\!\!\mathrm dT''\frac{v_w(T'')}{H(T'')}\right)^{\!3}.
\label{eq:Tp}
\end{align}
The percolation temperature is defined by $I(T_p)=0.34$, corresponding to $P(T_p)= 0.7$ \cite{Rintoul_1997}. Using $T_*=T_p$ in \eqref{eq:beta-tilde} yields the dimensionless inverse duration time.

\subsection{Strength Parameter \texorpdfstring{$\alpha$}{alpha}}
\label{sec:alpha}
Many analysis have used the MIT bag model to obtain the strength parameter $\alpha$ of the phase transition. As already mentioned in \cref{sec:Polyakov_model}, the bag model is not sufficient to precisely describe the confinement-deconfinement phase transition, and the Fuzzy Bag model \cite{Pisarski:2006yk} is required. In the bag model, the bag constant $\epsilon$ is used to parameterize the strength of the phase transition
\begin{align}
\alpha=\frac{\epsilon}{a_+T^4}\,.
\end{align}
The bag constant parameterizes the jump in both the pressure and energy density across the phase boundary
\begin{align}
\label{eq:bag_EoS}
p_+ &= \frac13 \,a_+ T^4 + \epsilon \,,
&
e_+ &= a_+ T^4 - \epsilon \,, \notag \\
p_- &= \frac13 \,a_- T^4 \,,
&
e_- &= a_- T^4 \,.
\end{align}
For work that defines the strength parameter beyond the bag model, see, e.g., \cite{Giese:2020rtr, Giese:2020znk}. Here, we define the strength parameter $\alpha$ from the trace of the energy-momentum tensor $\theta$
\begin{align}
\alpha=\frac{1}{3}\frac{\Delta\theta}{w_+}=\frac{1}{3}\frac{\Delta e\,-\,3\Delta p}{w_+}\,,
\end{align}
where $\Delta X= X^{(+)}-X^{(-)}$ for $X = (\theta$, $e$, $p$) and $(+)$ denotes the meta-stable phase (outside of the bubble) while $(-)$ denotes the stable phase (inside of the bubble). The enthalpy density $w_{\pm}$ is defined by
\begin{align}
w_{\pm}=\frac{\partial p}{\partial \ln T}\bigg\vert_{\pm}\,,
\label{eq:enthalpy}
\end{align}
which encodes the information of the number of relativistic degrees of freedom (d.o.f). It is intuitive to use the trace of the energy momentum tensor $\theta$ to quantify the strength of the phase transition $\alpha$. In the limiting case when $\theta=0$, the system possesses conformal symmetry and there is a smooth second-order phase transition occurring. $\theta$ is a quantity to measure the deviation from the conformal symmetry and thus also measures the deviation from the second-order phase transition. The larger $\theta$ is, the further away from the conformal symmetry and second-order phase transition and thus the stronger the first-order phase transition is. 

In the case of the confinement-deconfinement phase transition, $\alpha$ can be directly computed from the lattice results of $\Delta e$ and $\Delta p$ of \cite{Panero:2009tv}, see \cref{sec:fitting}. In our language of the PLM potential, we set the pressure and energy in the symmetry-broken phase to zero and measure energy and pressure relative to this phase, $e_-\sim p_-\sim 0$. Thus, $\alpha$ can be rewritten in terms of the $V_{\text{eff}}^{(+)}$
\begin{align}
\alpha=\frac{1}{3}\frac{4V_{\text{eff}}^{(+)}-T\frac{\partial V_{\text{eff}}^{(+)}}{\partial T}}{-T\frac{\partial V_{\text{eff}}^{(+)}}{\partial T}}\,,
\label{eq:alpha}
\end{align}
where we have used 
\begin{align}
p^{(+)}&=-V_{\text{eff}}^{(+)}\,,
&
e^{(+)}&=T\frac{\partial p^{(+)}}{\partial T}\,-\,p^{(+)}
\end{align}
as well as \eqref{eq:enthalpy}. Furthermore, at the percolation temperature (which is close to $T_c$), we always have $e_+\gg p_+$ \cite{Panero:2009tv}, leading to $\alpha \approx 1/3$. Note that our definition of $\alpha$ only depends on the degrees of freedom in the hidden sector. In other works \cite{Breitbach:2018ddu, Fairbairn:2019xog, Archer-Smith:2019gzq}, two different $\alpha$ have been introduced where one of them is denoted by $\alpha_\text{d}$, identical to the one defined in  \eqref{eq:alpha}, and the other is $\alpha_\text{tot} = \Delta\theta/3w_\text{tot}^{+}$, in which $w_\text{tot}^{+}$ is the total enthalpy including the visible and dark relativistic degrees of freedom. The parameter $\alpha_\text{d}$ is then used to compute the wall velocity and efficiency factors, while $\alpha_\text{tot}$ is used in the GW formula for the peak amplitude. To avoid the confusion, we only define a single $\alpha$ but take into account the dilution effect on the GW signals due to the presence of other degrees of freedom, see \cref{sec:spectrum} for more details.

\subsection{Bubble-wall velocity}
\label{sec:wall-velocity}
The bubble-wall velocity $v_w$ is another important parameter, which determines the strength of the GW signal. The bubble-wall velocity requires a detailed analysis of the forces that act on the bubble wall. The forces can be divided into two parts. The first force arises from the difference of the vacuum potential (pressure) between the confinement and deconfinement phases. This force accelerates the wall and causes the bubble to expand. The second force is the friction on the wall, which can be further divided into two kinds as discussed below \cite{Bodeker:2009qy, Bodeker:2017cim, Cai:2020djd, Baldes:2020kam}. For more recent work which calculates the bubble wall velocity beyond the leading-log approximation see e.g.~\cite{Wang:2020zlf}.\smallskip

\textbf{Direct Mass Change} The first kind of friction is due to the direct mass change of a particle when passing through the interface between the two phases (first proposed in \cite{Bodeker:2009qy}). The mass change results in a momentum change along the bubble moving direction, leading to a friction force on the bubble wall
\begin{align}
\frac{F_1}{A}=p_{\text{f}1}=\sum_a\Delta m_a^2\int\frac{\mathrm d^3p}{\left(2\pi\right)^3}\frac1{2E_{p,\text{dec},a}}f_a(p,\text{in})\,,
\label{eq:friction1}
\end{align}
where $F_1,\,A,\,p_\text{f}$ denotes respectively the friction force, the surface area and the pressure on the wall associated with the friction force. $\Delta m_a^2$ represent the mass square difference between the stable phase and meta-stable phase for the particle species $a$. $f_a(p,\text{in})$ is the distribution function for the incoming particles i.e.~in the deconfinement phase. In the phase transition from deconfinement to confinement, the gluons will confine to glueballs and become massive. A detailed estimate of this friction force relies on the estimate of the glueball mass of different numbers of colours. More importantly, \eqref{eq:friction1}, rigorously speaking, is derived from $1\rightarrow 1$ process (one incoming particle and an outgoing one) whereas the formation of glueballs from gluons is more complicated -- processes such as $2\rightarrow 1$ and $3\rightarrow 1$ may take place. In this case, a generalization of \eqref{eq:friction1} would be required. A more detailed study on the glueball formation is beyond the scope of this work and will be pursued in the future.
Nevertheless, there certainly exists friction in light of the direct mass change from gluons to glueballs.
\smallskip

\textbf{Particle Splitting} The second kind of friction (first discussed in \cite{Bodeker:2017cim}) is through the particle splitting (transition radiation process) where an incoming particle changes its momentum (along the bubble wall direction) through emitting another particle that exerts a friction force on the bubble wall. It was shown in \cite{Bodeker:2017cim} that in a large class of transition radiation processes such as $S\rightarrow V_TS,\,F\rightarrow V_TF,\,V\rightarrow V_TV$, (where $S,\,V,\,F,\,T$ denote respectively scalar, vector, fermion and transverse modes), the friction is given by
\begin{align}
\frac{F_2}{A}=\gamma\,p_{\text{f}2}\propto\gamma g^2\Delta mT^3\,,
\label{eq:friction2}
\end{align}
where $\gamma$ is the Lorentz factor since the friction scales with the incoming particle density and $g$ is the coupling of the involved interaction. $\Delta m$ represents the mass change of the particle at the interface, which implies that the friction resulting from the particle splitting process will always be associated with the above-mentioned friction of the direct mass change at the interface. In a weakly coupled theory, this second kind of friction is sub-leading compared with the previous one. However, in our strongly coupled system, this second friction can be equally important. 
\smallskip

\textbf{Wall Velocity} In summary, we can write the total pressure on the bubble wall as (see also \cite{Ellis:2019oqb})
\begin{align}
p_\text{tot}=\Delta V-p_{\text{f}1}-\gamma\,p_{\text{f}2}\,,
\end{align}
where $\Delta V$ denotes the pressure due to the difference of the vacuum potential between the confinement and deconfinement phases, which accelerates the wall. Assuming $\Delta V>p_{\text{f}1}$, we can obtain the equilibrium $\gamma$ (denoted as $\gamma_\text{eq}$ below) when the net force on the bubble wall becomes zero and the wall velocity (also $\gamma$) ceases to grow
\begin{align}
\gamma_\text{eq}=\frac{\Delta V-p_{\text{f}1}}{p_{\text{f}2}}\,.
\label{eq:gamma_eq}
\end{align}
Using \eqref{eq:gamma_eq}, we can obtain the terminal wall velocity.
\smallskip

\textbf{Relation to Energy Budget} 
In the end, it is important to consider the fraction $E_\text{wall}/E_\text{V}$ where $E_\text{wall}$ corresponds to the wall energy at the terminal velocity and $E_\text{V}$ is the total vacuum energy. This fraction describes how much of the total vacuum energy goes into accelerating the bubble wall. This part will eventually contribute to the GW signal from the bubble collisions. The remaining part of the energy $1-\frac{E_\text{wall}}{E_\text{V}}$ goes into the surrounding plasma and contributes to the generation of GWs via sound waves and turbulence, where typically the sound wave contribution dominates. In the case of the deconfinement phase transition, both friction terms $p_{\text{f}1}$ and $p_{\text{f}2}$ are non-perturbative due to the strong gauge coupling. Thus the main part of the energy will be stored in the plasma surrounding the bubble wall and, in consequence, we can focus on the GW production from sound waves. Due to the non-perturbative nature of the friction terms, it is highly challenging to determine them quantitatively. Instead, we treat the terminal bubble wall velocity as an input parameter and investigate the impact of different values.

\subsection{Thin-Wall Approximation}
\label{sec:thin-wall}
The advantage of the thin-wall approximation is that we can calculate analytically the decay rate of the false vacuum in terms of the latent heat and the surface tension. The latter are provided from lattice results as a function of the number of colours $N$ \cite{Panero:2009tv, Lucini:2005vg}. The thin-wall formula for the Euclidean action is shown in \cite{Linde:1981zj, Fuller:1987ue} and we briefly review it below. The three-dimensional Euclidean action is written as
\begin{align}
S_3=\frac{4\pi}{3}r_c^3\left(p_\text{de}-p_\text{co}\right)+4\pi\sigma r_c^2\,,
\label{eq:S_3}
\end{align}
where $p_\text{de}$ and $p_\text{co}$ denote respectively the pressure in the deconfinement and confinement phase, $\sigma$ is the surface tension of the nucleation bubble, and $r_c$ is the critical radius of the nucleation bubble defined by
\begin{align}
p_\text{co}-p_\text{de}=\frac{2\sigma}{r_c}\,.
\label{eq:surface_tension}
\end{align}
On the other hand, the difference of the pressure between the deconfinement and confinement phase is also linked to the latent heat $L$ via
\begin{align}
p_\text{co}-p_\text{de}&=L\eta\,,
&
\eta&=\frac{T_c-T}{T_c}\,.
\label{eq:Latent_Heat}
\end{align}
Finally, by using \eqref{eq:surface_tension} and \eqref{eq:Latent_Heat}, the three-dimensional Euclidean action \eqref{eq:S_3} can be written as a function of latent heat $L$ and surface tension $\sigma$
\begin{align}
S_3=\frac{16\pi}{3}\frac{\sigma^3}{L^2\eta^2}\, ,
\label{eq:S_3_Lattice}
\end{align}
where the latent heat from the lattice results \cite{Lucini:2005vg} is  
\begin{align}
\frac{L}{N^2}&=\left(\!\left(0.766(40)-\frac{0.34(1.60)}{N^2}\right)\,T_c\right)^{\!4},
&
N&> 3\,.
\label{eq:latent_heat_Nc2}
\end{align}
The lattice error on the $1/N^2$ coefficient bares the largest uncertainty and will eventually contribute the most to the uncertainty of the GW parameters, as we will see later.
The surface tension on the other hand can be either proportional to $N$ or $N^2$ due to indecisive lattice results. Intuitively, one may expect that the strength of the phase transition increases with $N$. The strength however depends on both $L$ and $\sigma$ as shown in \eqref{eq:S_3_Lattice}, where $L/N^2$ is related to the latent heat per d.o.f -- \eqref{eq:latent_heat_Nc2} becomes independent of $N$ for $N \gg 1$ -- and $\sigma\propto N$ or $N^2$. As a result, the strength of the GW only grows with $N$ if $\sigma\propto N^2$. As we shall see in the next section, \cref{sec:thin_vs_PLM}, the PLM fitting prefers the $\sigma\propto N$ case. Nonetheless, we discuss both scaling behaviours of the surface tension in the following.\smallskip

\textbf{$\sigma$ proportional to $N$:}
In this case, the lattice fitting function of the surface tension is \cite{Lucini:2005vg}
\begin{align}
\sigma&=T_c^3\,(0.118(3)N-0.333(9))\,,
&
N&\geq 3 \,.
\label{eq:surface_tension_Nc}
\end{align}
By implementing \eqref{eq:surface_tension_Nc} and \eqref{eq:latent_heat_Nc2} to the Euclidean action \eqref{eq:S_3_Lattice}, we obtain
\begin{align}
S_3&=\frac{16 \pi (0.118 N-0.333)^3}{3 N^4 (0.759\, -\frac{0.34}{N^2})^8}\frac{T_c^3}{(T_c-T)^2}\,,
&
 N&\geq 3\,.
 \label{eq:S3_Lattice_N}
\end{align}
This function has an interesting behaviour: for fixed temperature factor $\frac{T_c^3}{(T_c-T)^2}$, $S_3$ has a maximum at $N\sim 11$. In the large-$N$ limit, the Euclidean action behaves as $S_3 \sim 1/N$, which implies that the effective PLM potential scales as $V_\text{eff}(\ell)\sim N^2$, see \eqref{eq:Euclidean_Action} and \eqref{eq:bounce-solution}.

From \eqref{eq:S3_Lattice_N} together with \eqref{eq:Tp}, we determine the percolation temperature. As a rule of thumb, the phase transition occurs around $S_3/T\sim 150$ for $T_c$ in the GeV range. For other $T_c$, this criterion changes with logarithmically with $T_c$. We observe that the difference between percolation temperature $T_p$ and critical temperature $T_c$ denoted as $\delta T=\lvert T_p\,-\,T_c\rvert$ starts to increase from $N=3$ until it reaches a maximum at $N=11$ and then gradually decreases. As mentioned above, one might naively expect that the strength of the phase transition increases with $N$. Thus $\delta T$, which relates to the strength of the phase transition, should also increase with $N$. However, this pattern only corresponds to $\sigma\propto N^2$ case. It should be noted that $\delta T$ at $N=11$ is around $20$ times bigger than $\delta T$ at $N=3$. A bigger value of the temperature difference $\delta T$ implies a longer duration and a stronger first-order phase transition\footnote{
%It has been shown that models with Coleman-Weinberg type symmetry breaking are particularly easy to have supercooling and thus a strong first-order phase transition 
Similar features sometimes are shown in the case of supercooling with a strong first-order phase transition and a longer duration
\cite{Konstandin:2011dr,Sannino:2015wka,Brdar:2019qut,Ellis:2020awk,Chishtie:2020tze,Huang:2020bbe,Eichhorn:2020upj}.} and a stronger GW signal. Thus, we expect an increasing GW signal from $N=3$ to $N=8$ using the aforementioned method of the PLM effective potential if surface tension $\sigma$ is proportional to $N$. 

\begin{figure}[t!]
\includegraphics[width=\linewidth]{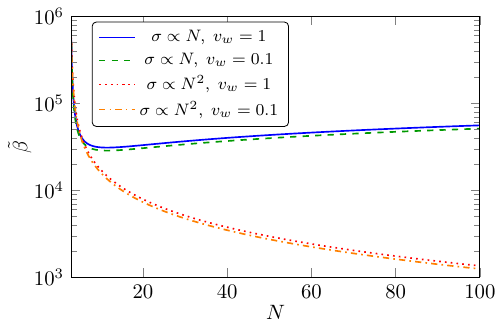}
\caption{Dimensionless inverse duration time of the phase transition as a function of the number of colours $N$ in the thin-wall approximation for different wall velocities $v_w$ and for different large $N$ behaviours of the surface tension $\sigma$.}
\label{fig:beta-TW}
\end{figure}

We can go one step further to derive the dimensionless inverse duration $\tilde \beta$. Using \eqref{eq:S3_Lattice_N}, \eqref{eq:Tp} and \eqref{beta}, we compute values of the dimensionless inverse duration $\tilde \beta$ which are shown in \cref{fig:beta-TW} with solid blue line and dashed green line. The $\tilde \beta$ shares exactly the inverse pattern as the more intuitive parameter $\delta T$ discussed above i.e.~$\tilde \beta$ first decreases to around $N=11$ and then increases with $N$. Since the gravitational wave peak amplitude is inversely proportional to $\tilde \beta$, it is consistent with the above discussion using $\delta T$.\smallskip

%%%%%
%%%%%

%%%%%
\begin{figure*}[t!]
	\includegraphics[width=\linewidth]{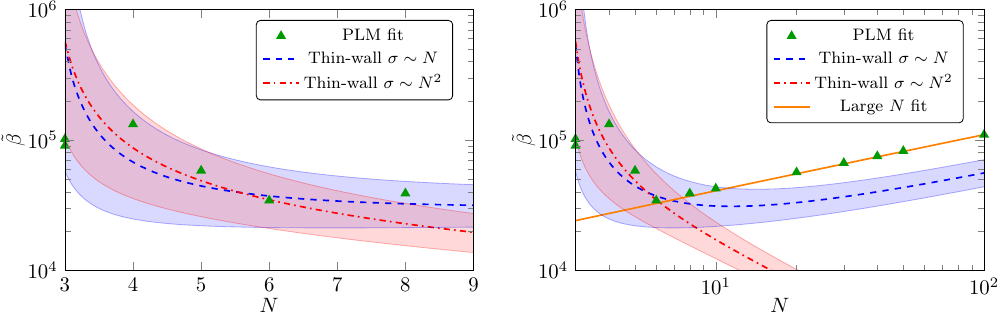}
	\caption{Inverse duration time of the phase transition as a function of $N$ in the effective PLM potential fit, in comparison to the thin-wall approximation. The bands for the thin-wall approximation include the lattice errors displayed in \eqref{eq:latent_heat_Nc2}, \eqref{eq:surface_tension_Nc}, and \eqref{eq:surface_tension_Nc_square}. The large-$N$ fit corresponds to the thin-wall approximation with $\sigma/T_c^3 = 0.075\, N$.}
	\label{fig:beta-PLM-vs-TW}
\end{figure*}
%%%%%

\textbf{$\sigma$ proportional to $N^2$:}
In this case, the lattice fitting function of surface tension is \cite{Lucini:2005vg}
\begin{align}
\sigma&=T_c^3\,(0.0138(3)N^2-0.104(3))\,,
&
N&\geq 3\,,
\label{eq:surface_tension_Nc_square}
\end{align}
while the latent heat is still following \eqref{eq:latent_heat_Nc2}. By substituting \eqref{eq:surface_tension_Nc_square} and \eqref{eq:latent_heat_Nc2} into the Euclidean action \eqref{eq:S_3_Lattice}, we obtain
\begin{align}
S_3&=\frac{16 \pi (0.0138 N^2-0.104)^3}{3 N^4 (0.766\, -\frac{0.34}{N^2})^8}\frac{T_c^3}{(T_c-T)^2}\,,
&
 N&\geq 3\,.
 \label{eq:S3_Lattice_N2}
\end{align}
This function has the following behaviour: For fixed temperature factor $\frac{T_c^3}{(T_c-T)^2}$, $S_3$ keeps increasing with the number of colours $N$. In the large-$N$ limit, the Euclidean action behaves as $S_3 \sim N^2$, which implies $V_\text{eff}(\ell)\sim 1/N^4$ as can be seen from \eqref{eq:Euclidean_Action} and \eqref{eq:bounce-solution}.

The pattern in this scenario is different from the previous case of $\sigma \propto N$. By using \eqref{eq:S3_Lattice_N2} and \eqref{eq:Tp}, we find that $\delta T=\lvert T_p\,-\,T_c\rvert$ monotonically increases with $N$. Thus, the larger $N$ the bigger $\delta T$, resulting in a stronger first-order phase transition and GW signals. Nonetheless, for both cases, $\sigma\sim N$ and $\sigma\sim N^2$, the thin-wall approximation gives consistent results for small $N$, i.e., $N\lesssim 7$. The ambiguity of the scaling behaviour of the surface tension can only be resolved in a strict sense by more accurate lattice results at large $N$. However, as we will show in the next section, the PLM fitting procedure seems so be only consistent with $\sigma\sim N$.

The dimensionless inverse duration time $\tilde \beta$ is again calculated by \eqref{eq:S3_Lattice_N2}, \eqref{eq:Tp} and \eqref{beta}, and the results are summarized in \cref{fig:beta-TW} with dashed red and orange lines. The $\tilde \beta$ shares exactly the inverse pattern as the more intuitive parameter $\delta T$ discussed above i.e.~$\tilde \beta$ keeps decreasing with $N$. Since the gravitational wave peak amplitude is inversely proportional to $\tilde \beta$, it is consistent with the above discussion using $\delta T$.

\subsection{Thin-Wall Approximation vs Fitting of PLM Potential}
\label{sec:thin_vs_PLM}
In this section, we compare the results of the dimensionless inverse duration $\tilde \beta$ between the two methods: the thin-wall approximation and the PLM potential fitting. We also comment on the wall thickness, which depends on the number of dark colours, and relates to the validity of the thin-wall approximation.

The comparison of $\tilde \beta$ is displayed in \cref{fig:beta-PLM-vs-TW}. The result of the fitted PLM potential, marked by green triangles, shows the following pattern: apart from $N=3$, $\tilde \beta$ first decreases with $N$ and then increases after $N=6$. Interestingly, this pattern qualitatively agrees with that of the thin-wall approximation with $\sigma\propto N$, although there the turning point is located around $N\sim 11$. For the thin-wall approximation, we include error bands due to the lattice error displayed in \eqref{eq:latent_heat_Nc2}, \eqref{eq:surface_tension_Nc}, and \eqref{eq:surface_tension_Nc_square}. The main error stems from the $N^2$ coefficient in the latent heat \eqref{eq:latent_heat_Nc2}. Note that we do not display the statistical uncertainties of $\tilde \beta$ associated with preferred regions of the $\chi^2$ fits in \cref{fig:beta-PLM-vs-TW} since they are small compared to the dot size (of the order of 10\%). However, the systematic uncertainties of the lattice data discussed in \cite{Panero:2009tv} have not been included in our fitting procedure. They may give rise to larger uncertainties on $\tilde \beta$. In the later computation of GW signals, we try to include those uncertainties by enhancing the statistical error by a generous factor of five.

On the right panel of \cref{fig:beta-PLM-vs-TW}, for points of $N>8$ without available lattice data, we assume that the energy and the pressure normalised to the SB limit become independent on $N$ in the large-$N$ limit. These assumptions are supported by lattice data for the pressure and energy \cite{Panero:2009tv}. This entails that $p\sim e \sim N^2$ in the large-$N$ limit and thus the effective PLM potential scales as $N^2$, $V_\text{eff}(\ell) \sim N^2$. In this case, the potential for $N>8$ can be obtained by a simple rescaling that of $N=8$, i.e., $V_\text{eff}(N) =N^2 V_\text{eff}(N=8)/8^2$. As discussed in the previous section, the scaling of the potential with $N^2$ corresponds to the scenario of $\sigma\propto N$ in the thin-wall approximation. 

We observe that $\tilde \beta$ from rescaled PLM potentials has a power-law behaviour as a function of $N$ (linear function in the log-log plot in \cref{fig:beta-PLM-vs-TW}). Intriguingly, the $N=6$ data point, which is obtained by using the direct lattice results rather than through rescaling, is in good agreement with the rescaling results. This seems to indicate that the information encoded in the lattice results for $N=6$ and $N=8$ favours the scenario of $\sigma\propto N$ rather than $\sigma\propto N^2$. Note that the blue curve in \cref{fig:beta-PLM-vs-TW} also becomes a linear function in the log-log plot for large $N$ but with a slightly smaller slope compared to that of the PLM fitting potentials. 

The fact that the PLM fitting favours $\sigma\propto N$ over $\sigma\propto N^2$ has a direct impact on the peak amplitude of the GW signal as discussed in the next two sections. It implies that the GW peak amplitude first increases (corresponding to a decreasing peak frequency) from $N=4$ to $N=6$ which has the lowest frequency and then gradually decreases~(while the peak frequency increases) with increasing $N$. On the other hand, for the case $\sigma\propto N^2$, which is not favoured by the PLM fitting potential, the signal becomes monotonically stronger with larger $N~(\geq 4)$.

\begin{figure}[t]
	\includegraphics[width=\linewidth]{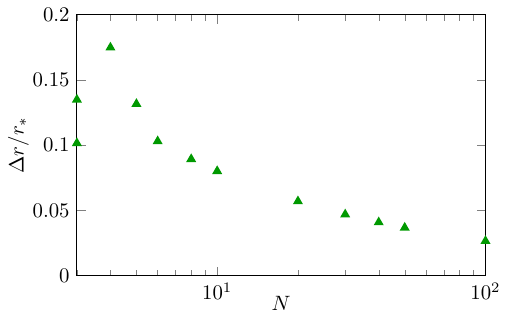}
	\caption{The ratio between the wall thickness $\Delta r$ and the bubble radius $r_*$ as a function of $N$.}
	\label{fig:wall}
\end{figure}

Before discussing the GW spectrum, we comment on the wall thickness. The wall thickness $\Delta r$ and the bubble radius $r_*$ can be directly computed from the instanton solution at the percolation temperature, see \cref{fig:instanton_solution}. We choose the wall thickness definition that the two wall boundaries are located $10\%$ away from the broken and unbroken Polyakov loop vacuum expectation values\footnote{Alternatively the wall thickness can be computed as the mass (second derivative of the PLM potential) at the confinement phase.}. We show the ratio of the wall thickness to the bubble radius in \cref{fig:wall} where the values for cases of $N>8$ are obtained via rescaling of the $SU(8)$ potential as discussed above. The wall is relatively thick for cases of $N=3,4,5$ and becomes thinner for $N \geq 6$. It continues to decrease in the large-$N$ limit. This is also consistent with what we have observed in \cref{fig:beta-PLM-vs-TW}; the results of PLM fitting potential are in agreement with or close to those of the thin-wall approximation for $N=6,\,8$ while more deviations between two methods are present for $N=3,\,4$.

\subsection{Gravitational-wave spectrum}
\label{sec:spectrum}
We briefly review the computation of the GW spectrum from the parameters $\alpha$, $\tilde \beta$, and $v_w$. In general, there are three contributions to the GW spectrum: collisions of bubble walls~\cite{Kosowsky:1991ua, Kosowsky:1992rz, Kosowsky:1992vn, Kamionkowski:1993fg, Caprini:2007xq, Huber:2008hg, Caprini:2009fx, Espinosa:2010hh, Weir:2016tov, Jinno:2016vai}, sound waves in the plasma after bubble collision~\cite{Hindmarsh:2013xza, Giblin:2013kea, Giblin:2014qia, Hindmarsh:2015qta, Hindmarsh:2017gnf} and magnetohydrodynamic turbulence in the plasma~\cite{Kosowsky:2001xp, Dolgov:2002ra, Caprini:2006jb, Gogoberidze:2007an, Kahniashvili:2008pe, Kahniashvili:2009mf, Caprini:2009yp, Kisslinger:2015hua}. As discussed in \cref{sec:wall-velocity}, in the case of the deconfinement phase transition, the contributions from sound waves are dominating and thus we focus on this contribution. Following \cite{Caprini:2015zlo, Caprini:2019egz, Wang:2020jrd}, the GW spectrum from sound waves is given by
\begin{align}
 h^2\Omega_\text{GW}(f)&= h^2\Omega^\text{peak}_\text{GW} \left(\frac{f}{f_\text{peak}}\right)^{\!3} \left[ \frac{4}{7}+\frac{3}{7}\left( \frac{f}{f_\text{peak}} \right)^{\!2}\right]^{-\frac{7}{2}}, 
 \label{eq:GWsignal}
\end{align}
with the peak frequency
\begin{align}
f_\text{peak}&\simeq 1.9\cdot 10^{-5}\,\text{Hz}\left(\frac{g_*}{100} \right)^{\!\frac{1}{6}}\left( \frac{T}{100\, \text{GeV}}\right) \left(\frac{\tilde \beta}{v_w} \right),
\label{eq:peak-f}
\end{align}
and the peak amplitude
\begin{align}\label{eq:peak-amp}
 h^2\Omega^\text{peak}_\text{GW} &\simeq 2.65\cdot 10^{-6}\left(\frac{v_w}{\tilde \beta}\right)\left( \frac{\kappa\, \alpha}{1+\alpha} \right)^{\!2}\left(\frac{100}{g_*}\right)^{\!\frac{1}{3}}\Omega_{SU(N)}^2\,.
\end{align}
Here, $h= H/(100 \text{km}/\text{s}/\text{Mpc})$ is the dimensionless Hubble parameter and $g_*$ is the effective number of relativistic degrees of freedom including the the SM degrees of freedom $g_{*,\text{SM}}=106.75$ and the dark sector ones $g_{*,SU(N)}=2(N^2-1)\cdot n$, where $n$ is the number of copies of dark sectors. The factor $\Omega_{SU(N)}^2$ accounts for the dilution of the GWs by the visible matter which does not participate in the phase transition. The factor reads   
\begin{align}
	\label{eq:dilution}
\Omega_{SU(N)} =\frac{\rho_{\text{rad},SU(N)}}{\rho_\text{rad,tot}}=\frac{g_{*,SU(N)} }{g_{*,SU(N)}+g_{*,\text{SM}}}\,,
\end{align}
where we again assumed that both sectors have the same temperature. In other works \cite{Breitbach:2018ddu,Fairbairn:2019xog,Archer-Smith:2019gzq}, two different strength parameters $\alpha_\text{tot}$ and $\alpha_\text{d}$ were introduced as discussed in \cref{sec:alpha}. In this case, the peak amplitude can be expressed in terms of these two quantities without 
involving the dilution factor:
\begin{align}
 h^2\Omega^\text{peak}_\text{GW} &\simeq 2.65\cdot 10^{-6}\left(\frac{v_w}{\tilde \beta}\right)\left( \frac{\kappa(\alpha_\text{d}) \alpha_\text{tot}}{1+\alpha_\text{tot}} \right)^{\!2}\left(\frac{100}{g_*}\right)^{\!\frac{1}{3}}\,.
\end{align}
Notice that the efficiency factor and the wall velocity depend on  $\alpha_\text{d}$ only.

%%%%%
%%%%%

In the last sections, we have detailed the computation of the parameters $\alpha$ and $\tilde \beta$, and argued that we use the wall velocity as a free input parameter. The last important ingredient is the efficiency factor $\kappa$, which describes the fraction of energy that is used to produce GWs. The efficiency factor is made up of the efficiency factor $\kappa_v$ \cite{Espinosa:2010hh} and an additional suppression factor due to the length of the sound-wave period~\cite{Ellis:2019oqb, Ellis:2020awk, Guo:2020grp}. In total, the efficiency factor $\kappa_\text{sw}$ is given by
\begin{align}
 \kappa_\text{sw}&= \sqrt{\tau_\text{sw}} \, \kappa_v\,.\label{eq:efficiency}
\end{align}
Note that we measure $\tau_\text{sw}$ in units of the Hubble time and thus it is dimensionless. We first discuss the contribution from $\kappa_v$ where we use the results from \cite{Espinosa:2010hh}. This efficiency factor depends on the wall velocity and the strength parameter. While it increases for larger $\alpha$, it typically has a maximum when the wall velocity assumes the Chapman-Jouguet detonation velocity $v_J$, which is given by
\begin{align}
	v_J = \frac{\sqrt{2\alpha/3 + \alpha^2} + \sqrt{1/3}}{1+\alpha} \,.
	\label{eq:detonationvel}
\end{align}
For the deconfinement phase transition where $\alpha \approx 1/3$, the detonation velocity takes the value $v_J = 0.866$. Due to the complicated dependence of the efficiency factor on the bubble-wall velocity, we simply display it for the wall velocities used here. In the next section, we test the impact of the wall velocity on the GW spectrum employing the values $v_w = (1,\, v_J,\,0.2)$. For $v_w=1$, $\kappa_v$ is given by
\begin{align}
\kappa_v(v_w=1)=\frac{\alpha}{0.73+0.083\sqrt{\alpha}+\alpha}\,,
\end{align}
which implies $\kappa_v \approx 0.3$ for $\alpha\approx 1/3$. At the Chapman-Jouguet detonation velocity $v_J$, the efficiency factor reads
\begin{align}
 \kappa_v(v_w=v_J)=\frac{\sqrt{\alpha}}{0.135 +\sqrt{0.98+\alpha}}\,,
 \label{eq:eff-vJ}
\end{align}
and for $\alpha\approx 1/3$ we have $\kappa_v \approx 0.45$. As expected, this value is larger that for $v_w=1$. For smaller wall velocities $v_w< c_s$, where $c_s=1/\sqrt{3}$ is the speed of sound, the efficiency factor decreases rapidly and the generation of GW from sound waves is suppressed \cite{Cutting:2019zws}. For example, for $v_w = 0.2$, we have
\begin{align}
\kappa_v(v_w=0.2)=\frac{6.9\alpha}{1.36-0.037\sqrt{\alpha}+\alpha}v_w^{6/5}\,,
\end{align}
which implies $\kappa_v \approx 0.19$ for $\alpha\approx 1/3$.

%%%%%
\begin{figure}[t]
	\includegraphics[width=\linewidth]{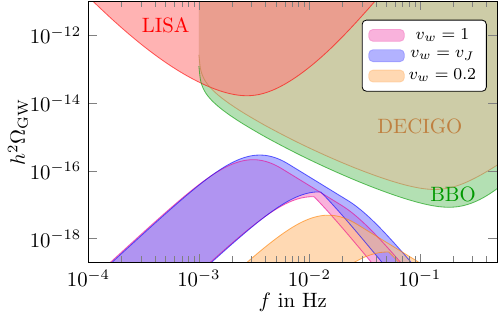}
	\caption{GW spectrum for the $SU(6)$ phase transition for different bubble-wall velocities, i.e., $v_w=1$, $v_w=0.2$, and $v_w=v_J$ the Chapman-Jouguet detonation velocity, see \eqref{eq:detonationvel}. The bands on the GW signal represent the small statistical lattice errors affecting the derivation of the $\alpha$ and $\tilde \beta$ parameters and include a further generous extra factor of five to account for hidden systematic errors. }
	\label{fig:GW1}
\end{figure}
%%%%%

%%%%%
\begin{figure}[b]
	\includegraphics[width=\linewidth]{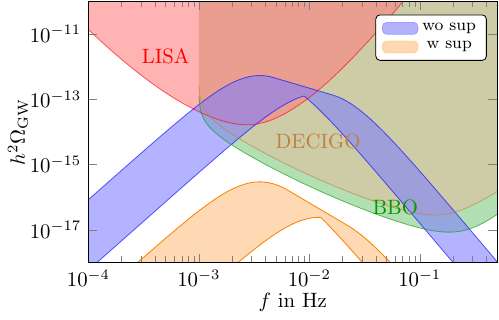} 
	\caption{Comparison of the $SU(6)$ GW spectrum with and without the suppression factor of \eqref{eq:suppression_1}.}
	\label{fig:GW2}
\end{figure}
%%%%%

%%%%%
\begin{figure}[t]
	\includegraphics[width=\linewidth]{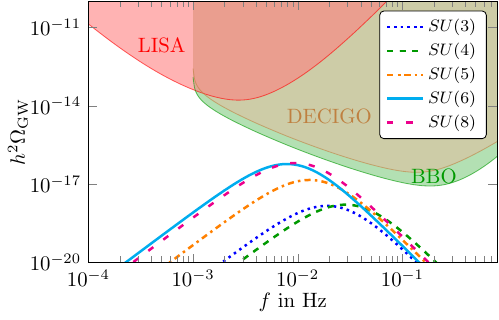}
	\caption{GW spectra from the $SU(N)$ deconfinement phase transition for different values of $N$. All spectra are plotted with the bubble-wall velocity set to the Chapman-Jouguet detonation velocity.}
	\label{fig:GW3}
\end{figure}
%%%%%

%%%%%
\begin{figure*}[t]
	\includegraphics[width=\linewidth]{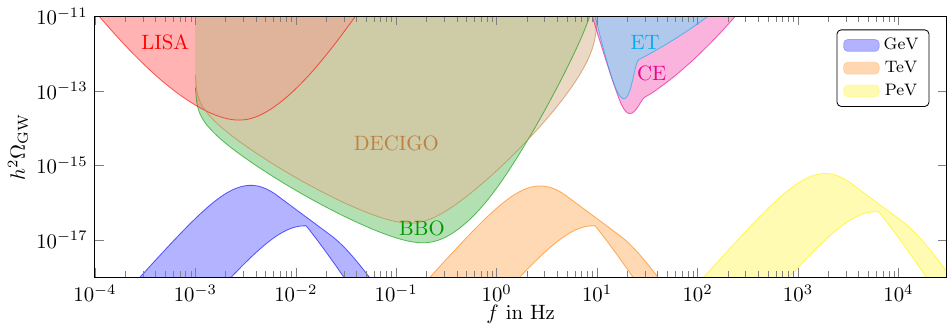}
	\caption{We display the GW spectrum of the $SU(6)$ phase transition for different confinement scales including $T_c = 1$\,GeV, 1\,TeV, and 1\,PeV. We compare it to the power-law integrated sensitivity curves of LISA, BBO, DECIGO, CE, and ET.}
	\label{fig:GW4}
\end{figure*}
%%%%%

The second contribution to the efficiency factor $\kappa_\text{sw}$ stems from a suppression due to the length of the sound-wave period $\tau_\text{sw}$, see \eqref{eq:efficiency}. In \cite{Ellis:2019oqb, Ellis:2020awk}, the length of the sound-wave period was given by
\begin{align}
\tau_\text{sw,1}= \text{min}\!\left[1, \frac{(8\pi)^{\frac 13} v_w}{ \tilde \beta \,\bar U_f}\right]. \label{eq:suppression_2}
\end{align}
The recent work \cite{Guo:2020grp} has analysed the length of the sound-wave period in an expanding universe and there the suppression was given by
\begin{align}
 \tau_\text{sw,2}=1-1/\sqrt{1+2\frac{(8\pi)^{\frac 13} v_w}{\tilde \beta \,\bar U_f}}\,. \label{eq:suppression_1}
\end{align}
They depend on the root-mean-square fluid velocity \cite{Hindmarsh:2015qta, Ellis:2019oqb}, which is given by
\begin{align}
\bar U_f^2 = \frac{3}{v_w(1+\alpha)}\int^{v_w}_{c_s}\!\mathrm d\xi \, \xi^2 \frac{v(\xi)^2}{1-v(\xi)^2}\simeq \frac{3}{4}\frac{\alpha}{1+\alpha}\kappa_v \,.
\end{align}
In our case, $\tilde \beta \gg 1$ and thus  \eqref{eq:suppression_2} and \eqref{eq:suppression_1} lead to almost identical suppression factors. For the subsequent analysis in the next section, we use \eqref{eq:suppression_1}.

An important quantity that determines the detectability of a GW signal at a given detector is the signal-to-noise-ratio (SNR) \cite{Allen:1997ad, Maggiore:1999vm} given by
\begin{align}
	\text{SNR} = \sqrt{\frac{T}{s} \int_{f_\text{min}}^{f_\text{max}}\mathrm df \left(\frac{ h^2 \Omega_\text{GW}}{h^2 \Omega_\text{det}} \right)^2}. 
	\label{eq:SNR}
\end{align}
Here, $h^2 \Omega_\text{GW}$ is the GW spectrum given by \eqref{eq:GWsignal}, $h^2 \Omega_\text{det}$ the sensitivity curve of the detector, and $T$ the observation time, for which we assume $T=3$\,years. We compute the SNR of the GW signals for the next generation of GW observatories which are LISA \cite{Audley:2017drz, Baker:2019nia, LISAdocument}, BBO \cite{Crowder:2005nr, Corbin:2005ny, Harry:2006fi,Thrane:2013oya, Yagi:2011wg}, DECIGO \cite{Seto:2001qf,Yagi:2011wg, Kawamura:2006up,Isoyama:2018rjb}, ET \cite{Punturo:2010zz, Hild:2010id, Sathyaprakash:2012jk, Maggiore:2019uih}, and CE \cite{Evans:2016mbw, Reitze:2019iox}\footnote{For an overview on challenges and opportunities of GW detection at large frequencies, see \cite{Aggarwal:2020olq}}. The sensitivity curves of these detectors are nicely summarised and provided in \cite{Schmitz:2020syl}. It is in general a difficult question, from which SRN onwards a GW signal will be detectable, a typical estimate being $\text{SNR}>1-10$. 
This issue is also linked to how well the astrophysical foreground such as gravitational radiation from inspiralling compact binaries is understood and can be subtracted from the signal, see, e.g., \cite{Cutler:2005qq, Pan:2019uyn}. Here, we make the optimistic assumption that a signal with $\text{SNR}>1$ is detectable.

\section{Results}
\label{sec:GravityWaves}
\begin{figure*}[t]
	\includegraphics[width=\linewidth]{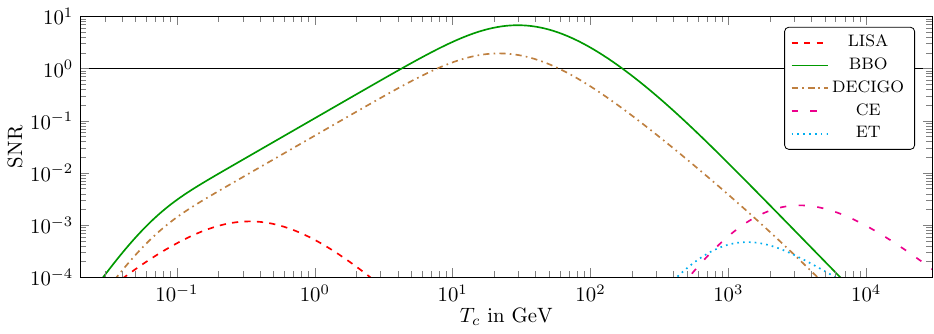}
	\caption{We display the SNR for the phase transition in a dark $SU(6)$ sector as a function of the confinement temperature $T_c$ from experiments of LISA, BBO, DECIGO, CE, and ET. We assume an observation time of three years.}
	\label{fig:SNR}
\end{figure*}

Here we discuss our main results on testing the dark confinement landscape using the next generation of GW observatories which are LISA, BBO, DECIGO, ET, and CE. We focus on the results obtained through the fitting of the effective PLM potential. The results from the thin-wall approximation are in qualitative agreement with those of the PLM potential fitting in particular in the case of the surface tension proportional to $N$ as discussed above\footnote{If the surface tension $\sigma$ is proportional to $N^2$, the GW signal increases with $N$. Thus for large-$N$ dark confinement phase transition, it will be even more strongly constrained by the future GW experiments compared with the $\sigma\propto N$ case. This may motivate the necessity for large-$N$ lattice simulations.}. The values of the dimensionless inverse duration $\tilde \beta$ for the different values of $N$ are displayed in \cref{fig:beta-PLM-vs-TW} for the wall velocity $v_w=1$. The dependence of $\tilde \beta$ on the wall velocity is only mild. The strength parameter takes values $\alpha\approx 1/3$, see \cref{sec:alpha}. From the PLM fitting, we obtain the statistical uncertainty on $\alpha$ and $\tilde \beta$, which we inflate by a generous extra factor of five to account for hidden systematic errors. We display this uncertainty with a band on the GW spectrum. For the GW experiments, we display in all figures the power-law integrated sensitivity curves, see e.g.~\cite{Alanne:2019bsm, Schmitz:2020syl}. The power-law integrated sensitivity curves can differ from standard sensitivity curves by several orders of magnitude. For the detectability of the GWs, we therefore refer strictly to the SNR, which is displayed in \cref{fig:SNR}. As input parameters for our computation, we have the wall velocity $v_w$, the confinement temperature $T_c$, the number of dark colours $N$, and the number of dark $SU(N)$ copies~$n$.

%%%%%
%%%%%

%%%%%
\begin{figure}[b]
	\includegraphics[width=\linewidth]{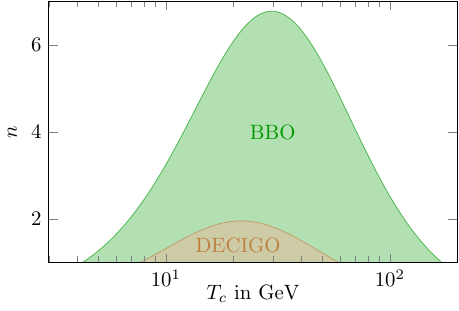}
	\caption{We display the exclusion curves of $n$ dark $SU(N)$ phase transitions from the experiments BBO, DECIGO as a function of the confinement temperature $T_c$. We assume an observation time of three years and that the signal is detectable for a signal-to-noise ratio $\text{SNR}>1$.}
	\label{fig:landscape}
\end{figure}
%%%%%

Let us start by discussing \cref{fig:GW1} where we show how different bubble wall velocities affect the GW spectrum using $SU(6)$ with $T_c=1$\,GeV as a testbed example. There are two competing effects at play here. The first is that the efficiency factor is maximal at the Chapman-Jouguet detonation velocity $v_J$, see \eqref{eq:eff-vJ}. The second is that the amplitude itself is proportional to the wall velocity, see \eqref{eq:peak-amp}. This means that at fixed $\tilde \beta$ and apart from the $v_J$ case, higher wall velocities tend to provide higher peak amplitudes and lower peak frequencies of GWs. 

In \cref{fig:GW2}, we compare the $SU(6)$ GW spectrum with and without the suppression factor given in \eqref{eq:suppression_1}. The suppression factor leads in our case typically to a suppression of $10^3-10^4$ or so. The suppression is significant for weak phase transitions and small for strong phase transitions. Excitingly the GW signal with suppression may still be detectable by BBO and DECIGO. Should the suppression of the GW signal due to the length of the sound-wave period be smaller than expected, then even LISA may be able to detect a signal from a deconfinement phase transition.

The dependence of the GW spectrum on the number of dark colours is shown in \cref{fig:GW3} for the values of $N=3,4,5,6,8$. All spectra are plotted with the bubble-wall velocity set to the Chapman-Jouguet detonation velocity and with $T_c=1$\,GeV. To make the figure concise, we do not show the error bands of the GW spectrum but only display the GW spectrum using the central values instead. From the plot, we learn that the peak amplitude of the induced GW signal is nearly independent of the number of colours for $N\ge 6$. That is due to the fact the strength depends on the jump in the entropy across the deconfinement phase transition per d.o.f.\ rather than the overall jump in entropy, which is inevitably proportional to $N^2$. This argument in principle also applies to small numbers of colors $N=3,4,5$, which is reflected in the overall mild dependence of $\tilde \beta$ on $N$ seen in \cref{fig:beta-PLM-vs-TW}. However, for small $N$, the GW signal is more strongly diluted by the d.o.f.~of the SM, see \eqref{eq:dilution}. The strength of the GW signal first increases (corresponding to decreasing peak frequency) starting from $N=4$ until reaching its maximal amplitude for $N=6$ (lowest frequency) and then the GW amplitude slowly decreases (with increasing frequency) when increasing $N$. This is in agreement with our expectations presented in \cref{sec:thin_vs_PLM} stemming from the dependence of the inverse duration time $\tilde \beta$ with respect to the dark colours $N$ shown in \cref{fig:beta-PLM-vs-TW}.

In \cref{fig:GW4}, we present how different confinement scales (including $T_c = 1$\,GeV, 1\,TeV, and 1\,PeV) affect the GW spectrum. As expected, a higher confinement scale leads to a higher GW peak frequency. On the other hand, the shape of the GW spectrum is independent of the confinement scale and also the peak amplitude depends only mildly on the confinement scale. Interestingly, BBO and DECIGO will test confinement phase transitions in the GeV range.

In \cref{fig:SNR}, we show the SNR of the phase transition of one dark $SU(6)$ sector for different GW detectors as a function of the confinement temperature $T_c$. Due to the minor dependence for $N\ge6$, as displayed in \cref{fig:GW3},
it is expected that cases of larger $N$ will feature similar SNRs shown here. Assuming that the signal is detectable for $\text{SNR}>1$, we find that BBO and DECIGO will test theories with a confinement scale within $1\,\text{GeV}\lesssim T_c \lesssim 100$\,GeV. Other GW detectors such as LISA, CE, and ET manage to achieve an SNR of $\mathcal{O}(10^{-3})$ in the GeV/TeV range. This analysis includes the suppression factor due to the short sound-wave period \eqref{eq:suppression_1}. The GW experiments will test a lager part of the landscape if the suppression factor is smaller than expected.

In \cref{fig:landscape}, we show how future GW observatories will constrain the dark deconfinement landscape. We make the assumptions that there are $n$ non-interacting copies of $SU(N)$ gauge theories and all of them undergo the phase transition at the same scale $T_c$.
The GW signals of these phase transitions add up linearly. However, due to the dilution of the GW signal over the non-participating degrees of freedom, the GW signal of each sector is suppressed by a factor of approximately $1/n^2$. Summing up the GW signal of all sectors leads to a total suppression of roughly $1/n$.
In other words, adding more independent sectors with the same scale of phase transition weakens the experimental constraints.
For each dark copies $n$ and $T_c$, we compute the SNR with respect to the future GW detectors BBO and DECIGO, see \eqref{eq:SNR}. We assume that the signal is detectable for $\text{SNR}>1$, and thus those theories will be tested in the future. Excitingly, BBO and DECIGO will cover the range of $1\,\text{GeV}\lesssim T_c \lesssim 100$\,GeV of this landscape and can maximally test four dark $SU(N)$ copies. The results in \cref{fig:landscape} again apply to scenarios of $N\ge6$. For $N<6$, the GW signal is slightly suppressed, see \cref{fig:GW3} -- the resulting SNR is smaller but the qualitative features of \cref{fig:landscape} still hold.

\section{Conclusions and Outlook}
In this work, we explored the landscape of the strongly coupled dark sectors composed of $n$-copy $SU(N)$ Yang-Mills confined theories coupled mainly gravitationally to our world. We employed state-of-the-art lattice results combined with effective field theory (PLM) approaches to investigate the GW signal arising from the dark deconfinement-confinement phase transitions in the early universe. As a comparison, we have also applied the analytic thin-wall approximation, which yields consistent results with those of the PLM approach. Our procedure is summarized in \cref{fig:work-scheme}.

We discovered that the strength of the GW signal only depends mildly on the number of colours $N$ for $N\ge6$. We find that the strength parameter of the phase transition is $\alpha \approx 1/3$, while the inverse duration time is $\tilde \beta = 10^4 - 10^5$ in units of the Hubble time.  Because of the fact that $n$ copies of a theory with the same confining scale need to share the same universe energy budget, we find that the next generation of gravitational waves observatories are sensitive only to a very small number of copies with confining scales from fraction to hundreds of GeVs as shown in \cref{fig:SNR}.

We consider our work a natural stepping stone towards the inclusion of matter fields in different representations of the gauge group. In particular, it is interesting to consider the interplay of the dark confinement phase transition and the one stemming from dark chiral symmetry breaking using the methodology of \cite{Mocsy:2003tr, Mocsy:2003qw} and the lattice results from \cite{Brower:2020mab}. Another avenue would be to extend the analysis beyond gauge-fermion theories to complete asymptotically free ones that feature composite dynamics including elementary scalars. These theories are well behaved at high energies while still featuring compositeness at low energies. Last but not the least it would be intriguing to investigate for these theories the GW imprint that could come from a dark symmetry, broken at arbitrary high temperatures as shown to exist in \cite{Bajc:2020gpa} following Weinberg's seminal work for UV incomplete theories~\cite{Weinberg:1974hy}. 
\bigskip

\noindent \textbf{Note added:}
shortly after this paper was submitted, we became aware of the nice complementary work \cite{Halverson:2020xpg}. They investigated the deconfinement phase transition using instead the Matrix Model rather than Polyakov Loop Model yielding compatible results with ours.

%%%%%%%%%%%%%%%%%%%%%%%%%%%%%%
\begin{acknowledgments}
We are grateful to M.~Panero for correspondence on the lattice results from \cite{Panero:2009tv}.
ZWW thanks Huan Yang for helpful discussions and MR acknowledges helpful discussions with M.~Hindmarsh, S.~Huber, and G.~Salinas.
This work is partially supported by the Danish National Research Foundation under the grant DNRF:90. 
WCH was supported by the Independent Research Fund Denmark, grant number DFF 6108-00623.
MR acknowledges support by the Science and Technology Research Council (STFC) under the Consolidated Grant ST/T00102X/1.
The authors would like to acknowledge that this work was performed using the \href{https://escience.sdu.dk/index.php/ucloud/}{UCloud} computing and storage resources, managed and supported by eScience center at SDU.
\end{acknowledgments}
 
\appendix

\bibliography{refs}

\end{document}